\documentclass{aastex6}
\turnoffedit

\begin{document}

\title{Chemical and physical parameters from X-ray high resolution spectra of
 \added{the Galactic nova}
V959 Mon}
\author{U. Peretz$^1$, M. Orio$^{2,3,}$\altaffilmark{4},
E Behar$^1$, A. Bianchini$^{3,5}$, J. Gallagher$^2$, T. Rauch$^6$, B. Tofflemire$^2$,
 P. Zemko$^5$}
\affil{$^1$Department of Physics, Technion, Haifa, Israel\\
$^2$Department of Astronomy, University of Wisconsin, 475 N. Charter Str.,
    Madison, WI 53706, USA\\
$^3$INAF, Osservatorio Astronomico di Padova, vicolo Osservatorio
 5, 35122 Padova, Italy\\
$^5$Dipartimento di Fisica e Astronomia, vicolo Osservatorio 3, Padova University, Italy\\
$^6$Institute for Astronomy and Astrophysics.
 Kepler Center for Astro and Particle Physics. Eberhard Karls
 University, Sand. 1., 72076 T\"ubingen, Germany
}
\altaffiltext{4}{marina.orio@oapd.inaf.it}


\begin{abstract}
Two observations of V959 Mon, done using the {\sl Chandra}
 X-ray gratings during the late outburst phases (2012 September and December),
 offer extraordinary
 insight into the physics and chemistry of this \edit2{Galactic} ONe nova.
 The X-ray flux was  1.7 $\times$ 10$^{-11}$ erg cm$^{-2}$ s$^{-1}$
and  \edit2{8.6} $\times$ 10$^{-12}$ erg cm$^{-2}$ s$^{-1}$, respectively
at the two epochs. The first result, coupled with electron density
 diagnostics and compared with published optical and ultraviolet observations,
 indicates that most likely in 2012 September the X-rays originated
 from a very small fraction of the ejecta, concentrated
 in very dense clumps.
 We obtained a fairly good fit to the September spectrum
 with a model of plasma in collisional
 ionization equilibrium (CIE) 
 with  two components; one at a temperature of 0.78 keV,
 associated with flat-topped and asymmetrical
 emission lines, blueshifted by $\simeq$710-930 km s$^{-1}$;
 the other at a temperature of 4.5 keV,
 mostly contributing to  the high-energy continuum.
 However, we cannot rule out 
a range of plasma temperatures between these two extremes;
we also modeled the spectrum as  a static cooling
flow, but the available models and the data quality are not adequate
yet to differentiate between the two-component fit and a
  smoothly varying temperature structure.
 In December, the central white dwarf (WD) became visible in X-rays. We estimate
 an effective temperature of $\simeq$680,000 K, consistent with a WD mass
 $\geq$1.1 M$_\odot$. The WD flux is modulated with the orbital period,
 indicating high inclination, and two quasi-periodic modulations
 with hour timescales were also observed.
 No hot plasma component with temperature above 0.5 keV was observed
 in December, and the blue-shifted component cooled to kT$\simeq$0.45 keV.
 Additionally, new emission lines due to a much cooler plasma appeared,
 which were not observed two months earlier.
 We estimate abundances and yields of elements in the nova wind
 that cannot be measured in the optical spectra
 and confirm the high  Ne abundance previously derived for this
 nova. We also find high abundance of Al, 230 times the
 solar value, consistently with the prediction that
  ONe novae contribute to at least 1/3rd of the Galactic yield of $^{26}$Al.
\end{abstract}

\keywords{X-rays: stars, stars: abundances, cataclysmic variables, novae: individual (V959 Mon)}



\section{Introduction} \label{sec:intro}
 Novae eruptions occur because of explosive CNO burning at the bottom
 of the envelope accreted onto WDs in binary systems. Dredged-up and mixed 
 nuclei of C, N and O
 act as catalysts accelerating the burning and causing the envelope to
 expand as the electron degeneracy is lifted.
 Convection brings beta decaying nuclei to the upper layers, heating
 the envelope and allowing mass ejection. The outflow eventually occurs, mostly,
 or only, through a radiation driven wind \citep[see][and references therein]{Kelly13}.
 Each outburst ejects mass
 into the interstellar medium (ISM), from a few 10$^{-7}$ M$_{_\odot}$ to
 a few 10$^{-4}$ M$_{_\odot}$, depending on the mass accretion
 rate and time necessary for the pressure at the bottom of the
 accreted layer to exceed the gravitational pressure. This pressure 
is mostly
 an inverse function of the WD mass \citep[][]{Starrfield12, Wolf13}. 
 
 Two types of novae are very important for Galactic chemistry.
 The first \edit2{type} are primordial novae \edit2{in} low metallicity binaries, that
should yield Ti and nucleosynthetic
 end-points around Cu-Zn \citep{Jose07}. The \edit2{second type are the} 
 more common oxygen-neon novae (ONe), attributed
 to  WDs of oxygen and neon. \edit2{These novae} inject a significant amount of
 dredged up Ne and Mg and many intermediate elements
 including Al, Ar, Cl, and F into the ISM.  Despite the rarity
 of ONe WDs, if \edit2{these WDs} accrete H-rich
 material, the outbursts are frequently repeated: there are probably 
 $\approx$15 a year \edit2{such eruptions}
 in the Galaxy, assuming an observed nova rate of 50 novae of all types each  year 
\citep[see][]{Shafter97, Shafter16}.  ONe novae play a very interesting role in Galactic chemistry,
and we have been able  to cast light on their high metallicities with the work we present in this article. 

 We analyse here in detail two {\sl Chandra} X-ray gratings' observations
 of the ONe nova V959 Mon, focusing also on the contribution of this nova
 to the Galactic chemistry. In Section 2 we review the known information about the
 nova. In Section 3 we describe the observations, and their analysis,
 including model fits for the ejecta and for the central WDs. 
 Section 4  focuses on the analysis of the ejecta, especially
 on their chemistry as derived from the ``X-ray harder'' 2012 September spectrum, 
 with a special attention for the problem of the yield of Al. Section 5
 describes the timing analysis of the 2012 December observation. Finally, in Section
6 we discuss our findings and derive conclusions.  
 
\section{V959 Mon, a nova discovered as a gamma-ray source}
 Nova Mon 2012 (V959 Mon), the best  studied 
ONe nova of recent years, was also
the first and only nova detected in
 gamma-rays before an optical observation. In fact, it exploded when it was angularly
 close to the Sun and could not be observed 
 at optical wavelengths until August 9, 2012 \citep{Fujikawa12}.
 Quite surprisingly, because only two
 other novae had been observed as gamma ray sources in the two
previous years, it was identified with a gamma ray transient discovered with
 Fermi on June 22, 2012 \citep[][]{Cheung12a, Cheung12b}. 
Unlike the previous two novae discovered as gamma ray sources,
 Nova Mon 2012 was a slow nova. \cite{Greimel12} 
  identified the quiescent counterpart, as
 an H$\alpha$ luminous, variable source of about 18th magnitude in
 the r filter. The time t$_3$ for a decay by 3  magnitudes, a useful
 quantity to classify novae, is not known. However, because in August of 2012 the spectrum   
 was already nebular, \citet{Munari13} inferred that t$_3$ 
 must have been quite shorter than 
 the 45 days  since the discovery of the Fermi transient. On the other hand,
 the light curve decay was smooth and slow after August 2012.
 The decrease in optical magnitude currently continues, after 
 a slow decline by only about 7 magnitudes in the
 first 33 months. \citet{Munari13}
 noticed that the only
 change in the rate of optical decline was a break around the  
 time the nova turned off as a supersoft X-ray source, in January 2013 
 \citep[][]{Page13}.

  The nova is most likely at a distance of about 1.4 kpc \citep[][]{Linford15,Munari13, Ribeiro13}
 although a larger distance, of 3.6 kpc,  has been proposed \citep[][]{Shore13}.
 It is effected by a large column density of neutral hydrogen
 with 5.9-7.4 $\times 10^{21}$ cm$^{-2}$ (measured towards the nova 
 direction according to the HEASARC nH on-line tool, see references therein).
 Optical spectra of the outburst show many Ne and Mg lines,
 indicating overabundance of these elements
 as typical
 for an outburst on an ONe WD \citep{Munari13, Shore13, Tarasova14}.
  A  modulation with a clear period of 7.1 hours was detected in different
 energy ranges, and it was identified
 with the orbital period \citep[][]{Munari13, Page13}.
 The nova was followed at optical
 and radio wavelengths as it returned to quiescence
 \citep[][]{Ribeiro13}. It was recently
 spatially resolved in the radio \citep[][]{Chomiuk14, Linford15},
  revealing that the mass outflow was not symmetric. There was a 
 fast, diffuse bipolar outflow and a dense, slowly moving torus of
material in the orbital plane of the binary. Synchrotron emission was
 detected at the interface between the equatorial and polar regions,
 which probably was also the origin of the gamma rays in the
 early post-outburst stage \citep[][]{Chomiuk14}.
 The same geometry of the ejecta was clearly inferred also by
 analysing the optical spectra \citep[][]{Munari13, Ribeiro13, Shore13}.
 \citet{Ribeiro13} measured  
 a peculiar asymmetric and double-peaked
 line profile, which they also modeled as a bipolar outflow
 with expansion velocity  reaching
 a maximum 2400$^{+300}_{200}$ km s$^{-1}$ at day
130 after the outburst, and inclination 82$^o\pm6^o$.
    \citet{Shore13} performed a similar analysis, concluding that the
 opening angle was of about 70$^o$, at inclination between 60$^o$ and
 80$^o$.  The optical lines observed by the above groups of authors have
 a broad double peaked profile, which is well
 pronounced for lines emitted in the outer region
 of the outflow, but is much less  evident for lines originating
 \edit2{in} the interior region of the ejecta.  

The only estimates of abundances
 in the optical spectra were done by \citet{Tarasova14}, who
used the relationship between the spectral
 line intensities and the ratio of the abundances of ions
 of N, O, Ne and Ar relative to hydrogen. This author assumed  an 
electron temperature of 10,000 K and with this assumption 
 measured the electron density, n$_e=10^{7}$ cm$^{-3}$ in September
 25, 2012 and about 1.5 $\times 10^{7}$ cm$^{-3}$ on October 20 and 24 of 2012,
 using the forbidden oxygen lines observed in the optical spectra.
 Later in the outburst,  Tarasova found   lower values of n$_e$,
  2.4  $\times 10^{6}$ cm$^{-3}$  and 2.9  $\times 10^{6}$ cm$^{-3}$  
 in March 18 and April 17 of 2013, respectively. The
 value of n$_{\rm e}$ derived by \citet{Shore13} using the same method
 and assuming the same electron temperature  was 
 3 $\times 10^{7}$ cm$^{-3}$ in November, 2012, so it is seems that there was a 
 rise in the first months, then the shell became diluted at
 late phases, as mass loss ceased. 
 \citet{Tarasova14} measured the abundances of several elements,
 although with two 
different methods her results differ by even 50\%. 
 \edit2{The} abundances \edit2{also} seem to have changed in time, with a clear increase
 in Ne. \citet{Tarasova14} attributed part
 of this increase to initial clumpiness
 and inhomogeneity of the shell, making
 the results less reliable in the first
 months. However, she attributed most of the increase 
 to a real effect:  \edit2{Ne and O abundances were actually increasing} because of
the ongoing dredge-up of these elements as the outburst proceeded.

 \edit2{Assuming} that the representative abundances for the nova shell
 should be the ones measured during 2013, when the ejecta homogenized and
 were better mixed,  
 Tarasova \edit2{suggested a reliable set of} abundances of several elements. 
  \citet[][]{Shore13} did not measure the abundances, but
 noticed that the spectra were extremely similar to those of a previous
 ONe nova, V1974 Cyg,  \edit2{except for Ne, which must have been more abundant
in V959 Mon than in V1974 Cyg}. 
Tarasova also 
estimated that the total ejected mass  was about 1.2$\times 10^{-4}$ M$_\odot$
 assuming a distance of 3.6 kpc, but radio imaging done by \citet{Linford15}
  implies that the true distance is only 1.4 kpc, reducing
 this value to $\simeq$1.8 $\times 10^{-5}$ M$_\odot$,
 in perfect agreement with the predictions
 of recent ONe nova models \citep{Bennett13}.

   V959 Mon was monitored in X-rays with {\sl Swift} and it immediately
 appeared as an X-ray source since
 2012 August 19, but it was not supersoft and luminous until 
 the following month of December. The  {\sl Chandra} high resolution spectrum obtained
 in September showed a bremsstrahlung continuum 
 and many emission lines of several intermediate
 atomic-number elements. The strongest 
lines were H-like MgXII at 8.42 \AA \ and Si XIV at 6.18 \AA \ \citep[][]{Ness12}.
\added{This spectrum is particularly interesting
because  X-ray grating measurements of novae
 emission lines in the 1.7-10 \AA \ range
 have been rare, while several atomic transitions that are specific of ONe
 novae occur in this range. 
 A {\sl Chandra} emission line spectrum was obtained for 
V382 Vel, a known ONe nova,
 but only with the LETG grating, which is
 less sensitive in this range, and at quite a late outburst
 phase: only Mg lines at $\simeq$8.4 and $\simeq$9.2
 \AA \ were clearly measurable shortwards of 10 \AA \ 
\citep[][]{Ness05}. Two recurrent novae 
 with massive WDs of unknown composition were observed
 using gratings while they were relatively hard X-ray sources.
RS Oph  did not show  very strong lines of Mg and Al,
\citep[][]{Nelson08, Ness09}, consistently with the ultraviolet spectra of
 a previous outburst \citep{Contini95}.
 V745 Sco was observed as a hard X-ray source with
 NuSTAR \citep[][]{Orio15} and shortly later 
 it was observed it with the {\sl Chandra} HETG. \citet[][]{Drake16} 
 found that the spectrum 
 could be fitted with solar abundances. These
two novae host a red giant, thus it   is 
 possible that the ejecta were abundantly mixed with the
 the red giant wind, making it difficult
 to establish the nature of the secondary.
 In short, this V959 Mon spectrum is unique, because
 no other confirmed ONe nova
 was observed with X-ray gratings when the ejecta, or at least
 a portion of them, were still so hot to produce
 the emission lines we measured for this nova.
It is very likely, of course, that also other novae underwent a phase
 in which an X-ray spectrum of emission lines in the 1.7-10 \AA \ range could have
 been observed and measured, but the
X-ray grating observations were not scheduled timely enough.
We will show in this paper that the  {\sl Chandra} exposures of V959 Mon,
 and especially this first one, indicate
a new avenue to measuring yields of ONe novae.
}

 Once the X-ray spectrum
 exhibited a supersoft luminous component \citep[][]{Nelson12} we proposed 
 a second grating observation in December
 with {\sl Chandra} and the low energy transmission grating (LETG) coupled
 with the HRC camera. Initial results, indicating a WD continuum
 with overimposed emission lines originating in the ejecta, were
 announced in ATel 4633 \citep[][]{Orio12}. 
   In this article we revisit and analyse both {\sl Chandra} observations 
 in detail, focusing especially on the contribution of this nova
 to the Galactic chemistry. 

\section{The observations}
 \begin{figure}[t]
     \centering
 	\includegraphics[width=1.0\linewidth]{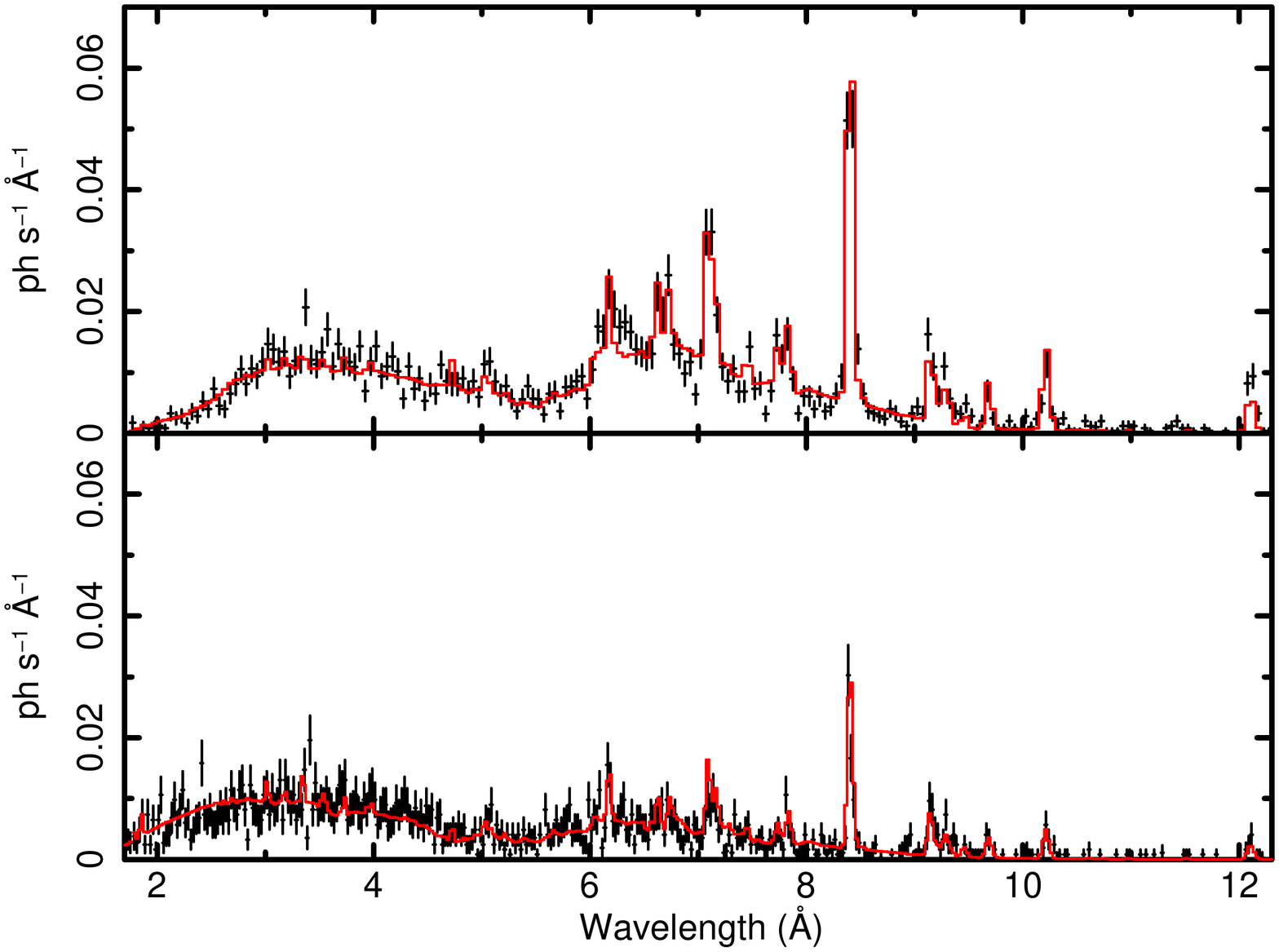}
 	\caption{The MEG spectrum (top
 panel) and the HEG spectrum (lower panel) measured with the {\sl Chandra} HETG on 
 		September 12, both binned with 10
 counts per bin, and the fit with the model
 shown in Tables \ref{table:params} and \ref{table:abundances}.}
 	\label{fig:spectra}
 \end{figure}
\begin{figure}[t]
        \centering
        \includegraphics[width=\linewidth]{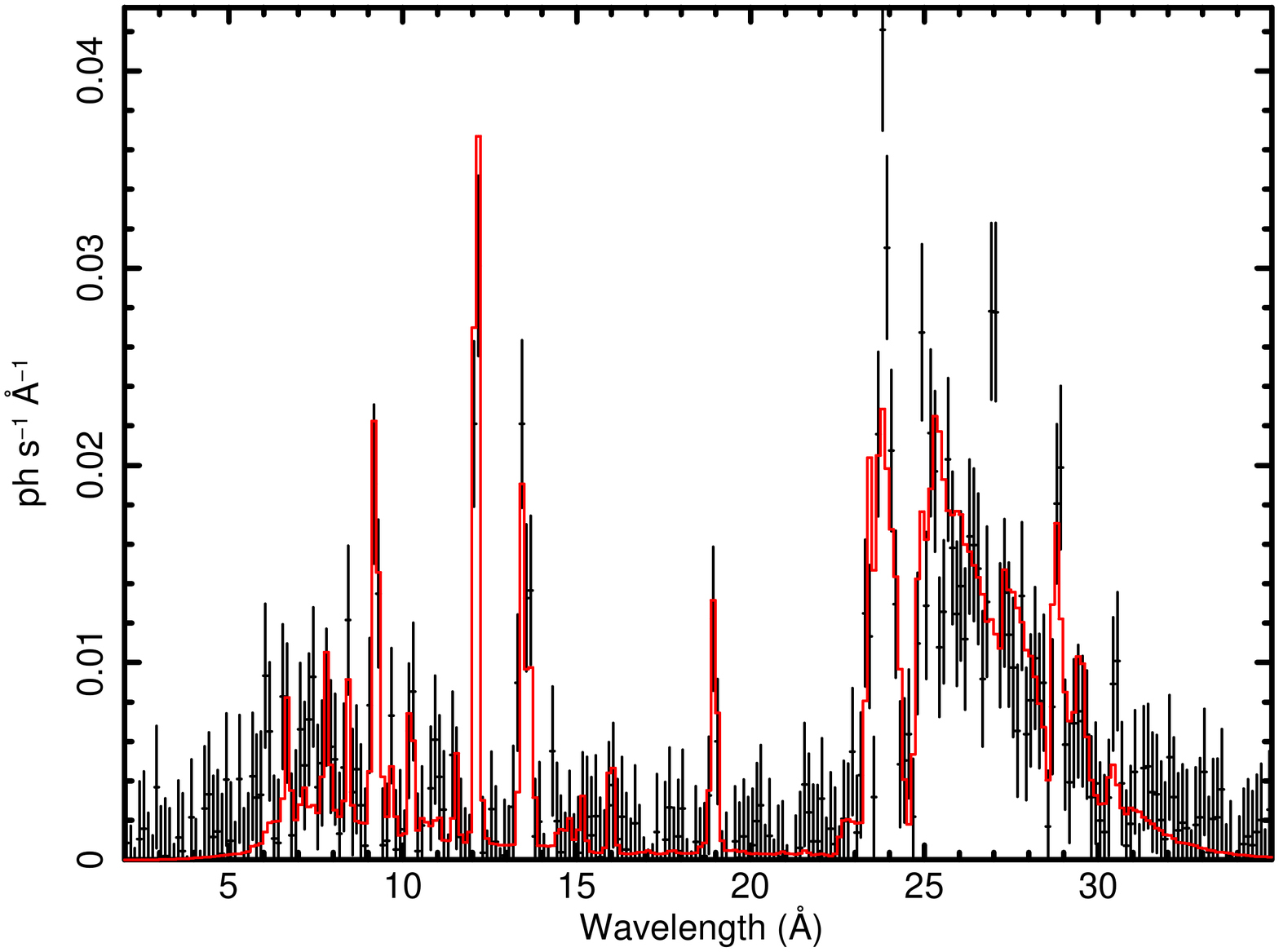}
        \caption{LETG spectrum
 measured in September of 2012 (in black). For visualizing
 purposes the spectrum has been binned
 with 10 counts per bin. The red line shows
 the fit with the composite model of two
 thermal plasma components in CIE and an atmospheric model.
The parameters of the fit
        are given in Tables \ref{table:params} and \ref{table:abundances}.}
        \label{fig:atmo}
\end{figure}

The first X-ray spectrum we analyzed, obtained from
 the {\sl Chandra} archive tgcat and shown in Fig.1, was taken
 on September 12, 2012, day 83 of the outburst, with the ACIS-S camera of 
{\sl Chandra} and the
 High Energy Transmission Gratings (HETG), following a TOO proposal by Ness
 et al. (2012). Both the medium
 energy MEG and the high energy HEG were used, with a
 respective absolute wavelength accuracy
 of 0.0006\AA\ and 0.011\AA.
The exposure time was 25 ks, with a measured count rate 
$0.18396\pm0.00280$ cts s$^{-1}$ in the zeroth
 order ACIS-S camera. Half of the incident radiation 
is dispersed to the gratings in this observation mode, and 
the count rates were
 0.04522$\pm$0.00097 cts s$^{-1}$ in the HEG 
 summed first orders (energy range 0.8-10 keV), 
 and 0.08035$\pm$0.00129  cts s$^{-1}$ in the MEG summed first orders
 (0.4-10 keV). 
\explain{A paragraph here has been moved to the previous section,
 as suggested by the referee.} 

 The second observation was proposed by us and carried out
 on December 03, 2012 (day 165) with the HRC-S camera and the low energy transmission
 grating (LETG), with wavelength accuracy of 0.04\AA\ \citep[][]{Orio12}. 
The spectrum is shown in Fig. 2.
 As in the previous observation, the exposure was
 a little over 25 ks long (25120 seconds).
The measured count rate was $0.13648\pm0.00240$ cts s$^{-1}$ with the
 HRC-s camera, and 0.2103$\pm$0.0161 with the LETG
 summed first orders.
 There is 
 almost no signal in the HETG spectra longwards of 15 \AA, and in the 
LETG spectrum longwards of 35 \AA.  We know
 that {\sl Swift} X-ray telescope observations clearly show that there
 was no soft X-ray flux in September 2012 \citep[][]{Nelson12, Page13}.

\subsection{The spectrum observed in September 2012}
\begin{figure}
 	\centering
 	\includegraphics[width=0.7\linewidth]{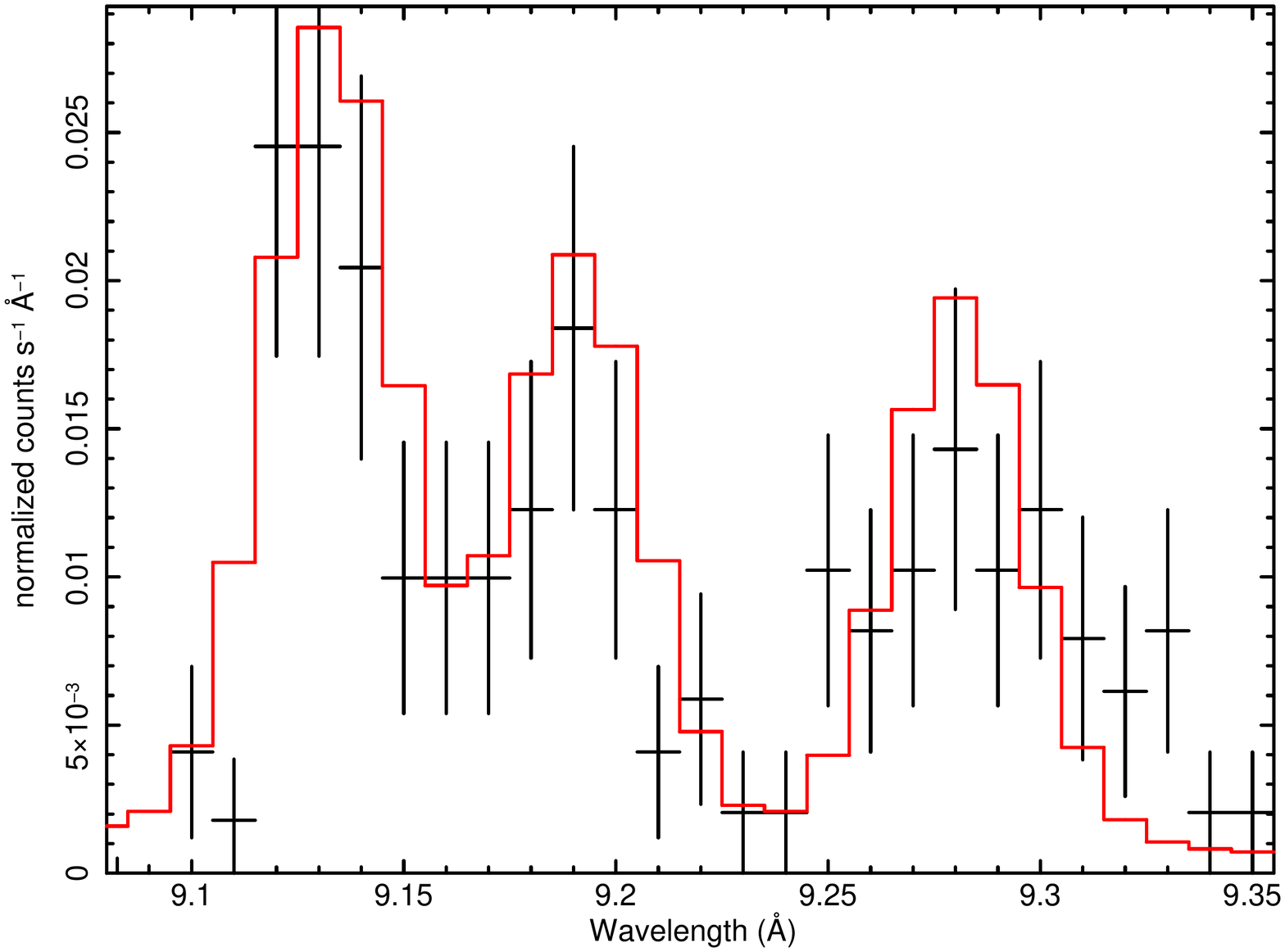}
        \includegraphics[width=0.7\linewidth]{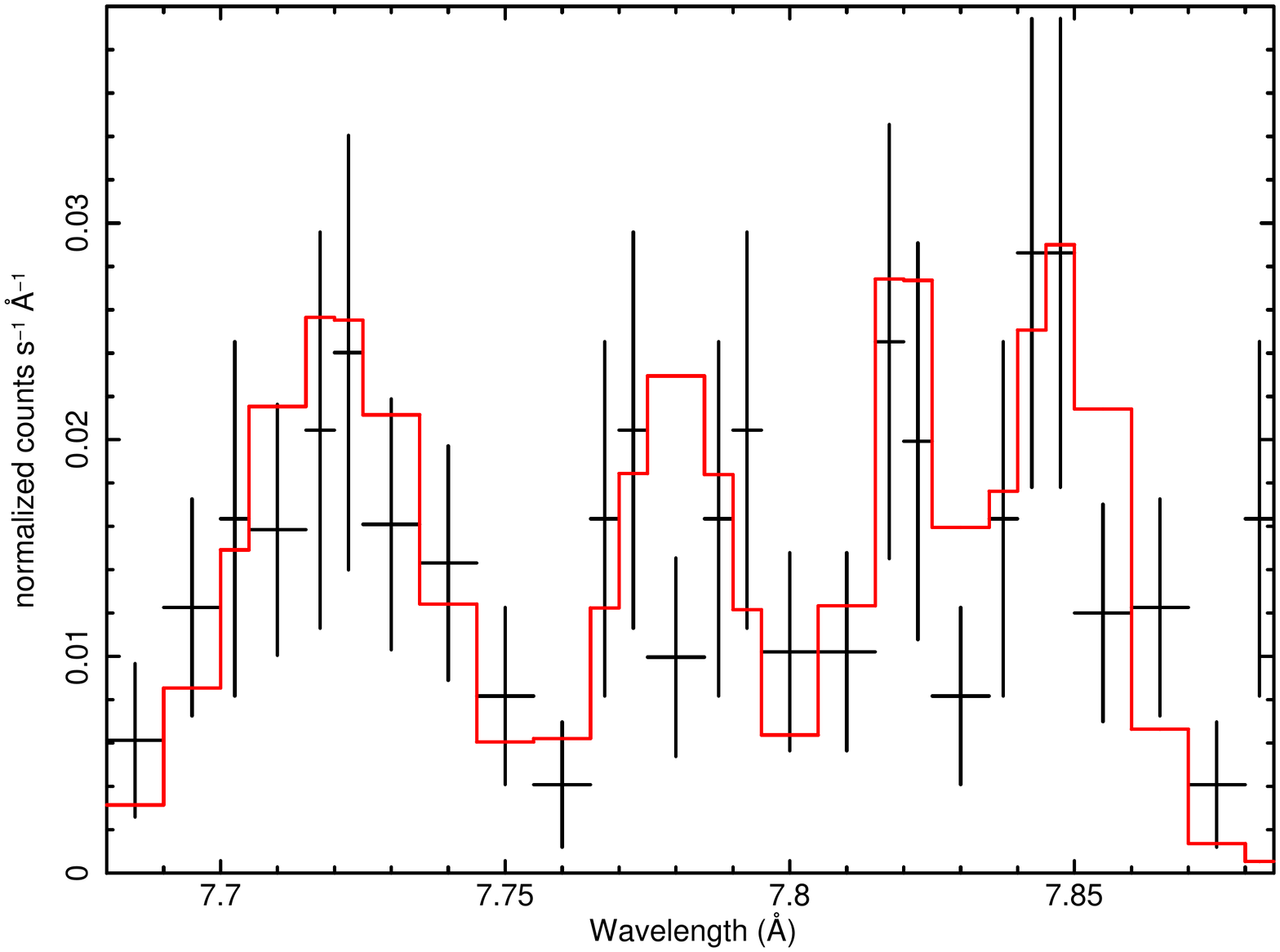}l
 	\caption{\explain{Figure
 added here.} The Mg XI He-like triplet and the Al XII He-like triplet
 with the Mg XI line at 7.85 \AA, measured
 		with the MEG and binned with 2 counts
 per bin (black data points). Only
 in order to guide the eye, we traced in red a possible fit with 
 Gaussians having the same width  (as calculated also 
 in the ``global'' model), but slightly
 different blueshift (however, see text for how the flux
 in the single lines was computed,
 without binning and independently of modeling).}
 	\label{fig:mgtriplet}
 \end{figure}

\begin{figure}
        \centering
        \includegraphics[width=0.5\linewidth, angle=90]{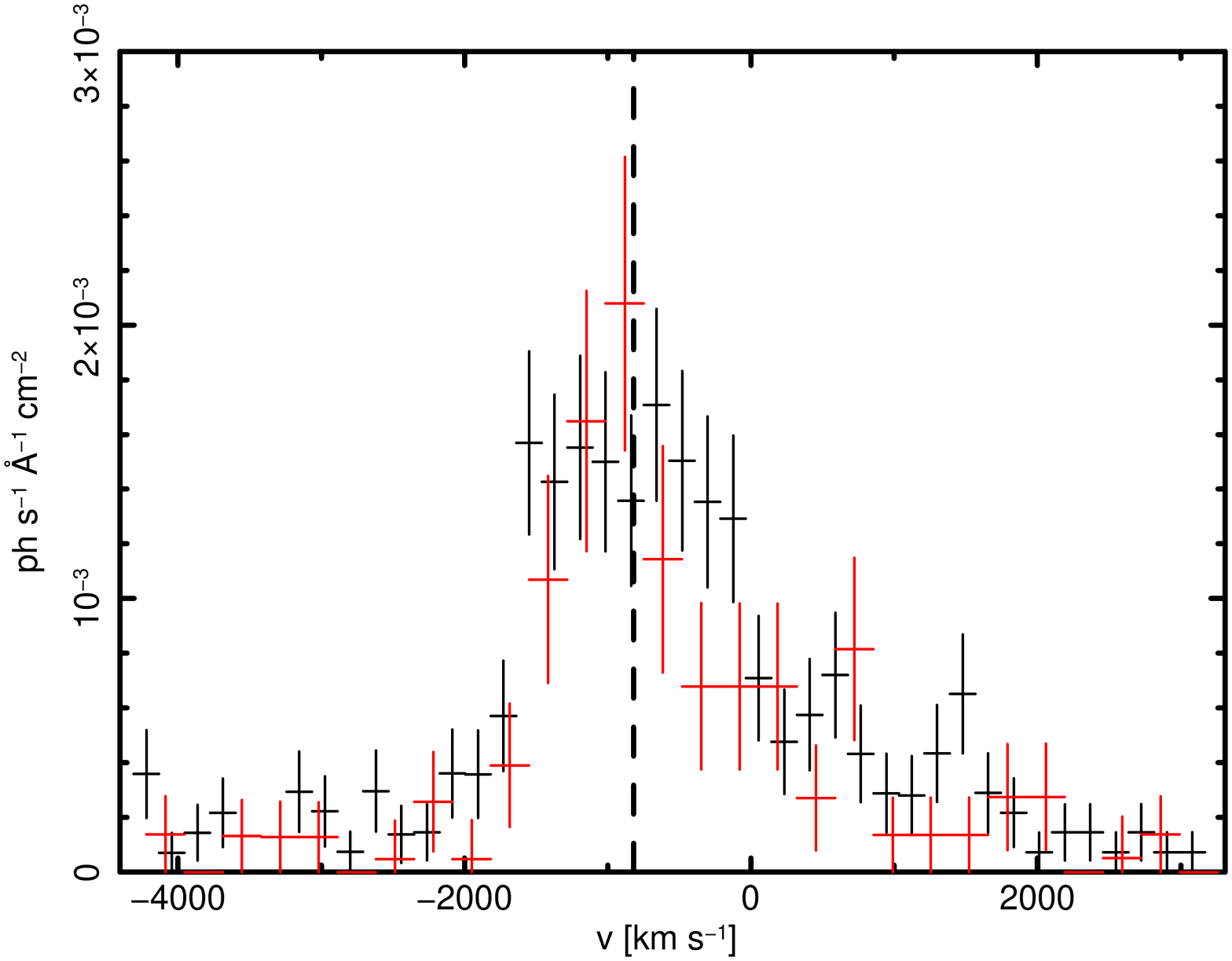}
        \includegraphics[width=0.8\linewidth]{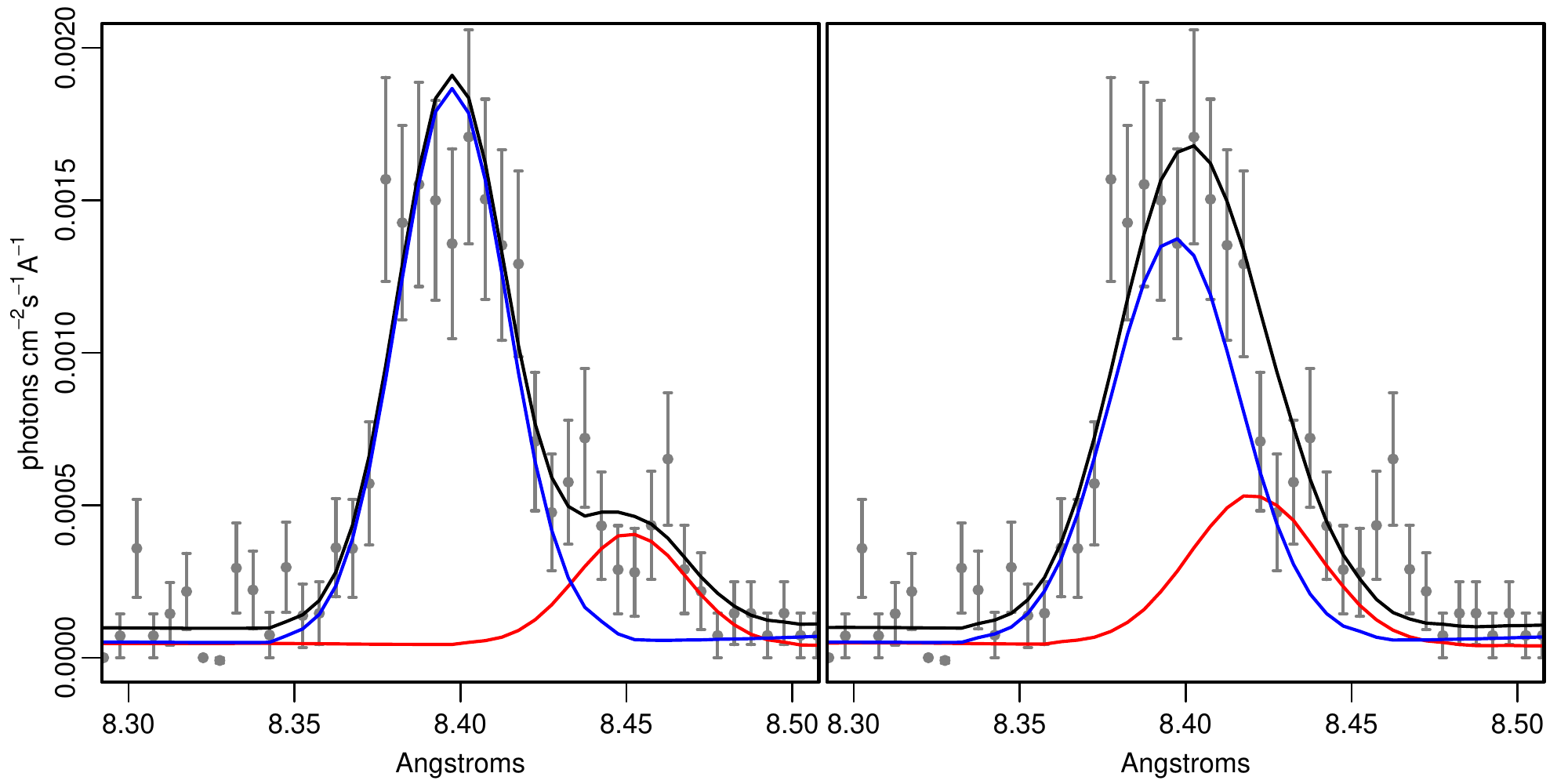}
 	\caption{\small In the top panel the    
 the Mg XII H-like line (an unresolved doublet with contribution
 		of both Ly$\alpha_1$ $\lambda$8.4192 and
  Ly$\alpha_2$ $\lambda$8.42461) is plotted in velocity space. The
 MEG data are in black and the HEG ones in red. The dashed line marks the line center
 of the less hot component, which is clearly blueshifted.
The lower  right panel shows the same line profile, with the MEG
 flux as a function of wavelength, 
as modeled in Table \ref{table:params}:
   the black solid line is the convolution of the two components of
 the model, of which the one at 0.78 keV is  plotted in blue and the
 one at 4.5 keV is in red. The right top panel shows
the same line and two CIE components model fit, obtained 
 allowing the velocity of the hotter component to vary 
and resulting in a redshift \replaced{by of}{of} the hot
 component by $\simeq$500 km s$^{-1}$. This fit
 is equally statistically significant for the overall spectrum, 
 and the other parameters it returns are only slightly different. 
}
 	\label{fig:mg-vel}
 \end{figure}

The HETG spectra, shown in   Fig. \ref{fig:spectra}, have low signal to noise, but
 we detect He-like triplets of several elements.
In the Si XIII, S XIV, Al  XII and Mg XI triplets (the latter shown
 in Fig. \ref{fig:mgtriplet}) the  resonance line  appears stronger than
 the inter-combination and forbidden lines (see Table \ref{table:fluxes}), although the
 intercombination line is not clearly resolved for Si and S. 
 Depending on the plasma temperature
 and specific element, the so called G ratio, G=${(f+i)}/r$, indicates whether
 the plasma is in CIE. 
 Generally G$>$4 indicates
 a contribution of photoionization \citep[][]{Porquet01, Bautista00},
 as long as this diagnostic is used in a regime where the forbidden
 line is not sensitive to the density, that is, extremely high densities. 
 With the method described below,
 we measured G$\simeq$2 for Mg, G$\simeq$1.5 for Al, although in the case 
 of Al there was a partial overlap of the f line with 
 a line of Mg XI \added{see Fig.
\ref{fig:mgtriplet}}. \added{Even taking into account
the errors in the measurement, as we see in Fig. \ref{fig:mgtriplet} 
 the {\it r} line is stronger than the other two of the triplet,
 so the G ratio value is much smaller than a value of 4.}
 These diagnostics points towards a plasma in CIE. 
The emission lines are \added{blueshifted and}
broadened, probably because
 the emission region was extended and \deleted{there were} not \added{in the} 
 collimated
 bipolar outflows inferred in the optical spectrum \deleted{and are blueshifted}. 
The profile of the lines, shown in Figs. \ref{fig:mgtriplet} and
 \ref{fig:mg-vel},
is asymmetrical and somewhat skewed towards the blue side. 
 Both the blue shift and asymmetry  may be due 
to intrinsic absorption in the nova shell, that erodes the 
part of the line formed in receding material on the far side. 
 The H-like lines have at least one additional, non-blueshifted component
 of lower luminosity, due to hotter plasma that can only contribute
 to the H-like lines because it is almost fully ionized (we will return to this
 when describing the global fit to the model). 

In Table \ref{table:fluxes} we give the measured fluxes for a few
 lines that could be relatively well
 isolated and were measured with good S/N. Since a Gaussian was not a good fit,
 we examined the spectrum to define
 the width of the line at the base and the continuum
 level, and \added{we added all
 the data points in order to calculate} the flux above the continuum. The error on the observed
 flux  was computed as the sum of the squares of the errors of
 the single data points. 
 The unabsorbed flux (or flux at the source) was calculated by assuming the
 value of the N(H) derived from a two-component global fit
 described below, with its additional uncertainty
 that propagates in the error. \added{In the table we also included the flux
 of some lines that partially overlap, like the {\it} r and {\it i} line of Mg XI
 in Fig. \ref{fig:mgtriplet}, calculated by simply truncating the
 summatory at the minimum point between the lines (e.g. 0.16 \AA \ in
  Fig. \ref{fig:mgtriplet}. This introduces an additional error
of the order of 10\%  in the
 flux calculation for such lines,  that is
 difficult to evaluate
 exactly and could not be included in Table \ref{table:fluxes}. However,
 we are mostly interested in flux ratios as diagnostics, and the uncertainty introduced by
 overlaps in the wings of  
 the lines' profiles
 does not imply additional uncertainty in the ratios
 as long as the lines have the same profile, as it is reasonable to assume. especially
 for lines of the same triplet.} 

We extracted the spectra \added{with the
 CIAO software version 4.8.1 
 and the calibration package CALDB version 4.7.1,
 then} we modeled \deleted{both} HEG and MEG spectra simultaneously with the 
 APEC code in XSPEC version 12.9 \citep[][]{Smith01}; more specifically,
 we used BVVAPEC. \added{We present the results with
 errors corresponding to the \edit2{90\% confidence 
 intervals}.}
 Consistently with the profile of the H-like lines and with the excess flux
 at short wavelengths,  we 
 obtained the best fit with two plasma components at different
 temperatures. A hot component explains the flux
 at the shorter wavelengths well, short-wards of $\simeq$5 \AA, 
 and a cooler
 component explains most of the flux in the emission lines.  
 Fig. \ref{fig:mg-vel} shows the example of the Mg  H-like emission line and how it is 
 fitted with the global \replaced{fit}{model} with two components. 
 The cooler plasma component has a much higher flux in the range 
 where most lines are observed. 
Because no emission lines with significant S/N appear in the lower wavelengths, 
constraining the velocity of the hot component is not feasible. Depending on
the exact parameterization of line widths, the fit converges to either positive or 
negative velocity. 
Since the resulting abundances do not depend on this velocity, we present a
 fit in which we set the hot component velocity to zero.
 The amount of blueshift due to the cooler component seems
 to vary for each line, from $\simeq$800 km s$^{-1}$ to
 1250 km s$^{-1}$, but all are consistent within the \edit2{90\% confidence level} error of the model.
 In the \replaced{lower}{upper}
 panel of Fig. \ref{fig:mg-vel} the same Mg line is plotted in velocity space with the model
 best fit blueshift marked. In our best fit, 
 the broadening parameter is tied in the two components (for lack of any better
 constraint), and the lines, which are all clearly broader than 
 the instrumental width, 
 have a fixed velocity at half maximum
 of \edit2{676} km s$^{-1}$, again visible in Fig. \ref{fig:mg-vel}. We obtained
 a good fit constraining also 
 the additional, hotter and  non-blueshifted  component of the H-like
 lines to have the same width.   
\added{Since we used unbinned data for the fit, we used C-statistics \citep{Cash79},
best suited for fits with a small number of counts per bin, rather than
$\chi^2$ statistics.} We found a Cstat parameter
 $c=5862.51$ and 6346 degrees of freedom.

 \begin{table}
 \small
\caption{Flux of prominent emission lines measured with the MEG. 
  The measurement method is described
 in the text, and the \replaced{absorbed}{unabsorbed} flux has been evaluated
 by assuming the value of N(H) of the global models (Table \ref{table:params}). 
 \added{The statistical error does not include the
 additional error uncertainty due to lines' overlap (however, as discussed in the text,
this does not significantly effect the line ratios). \explain{note that the
 flux of the LETG fit has changed, see Letter.}).}
 } 
\label{table:fluxes}
 \centering
 \begin{tabular}{lllll}
 \hline\hline
 Ion & Rest wavelength & \added{Observed wavelength} &  Absorbed Flux                       & Unabsorbed flux \\
     &       \AA     & \AA               & ph cm$^{-2}$ s$^{-1} \times 10^{-5}$ & ph cm$^{-2}$ s$^{-1} \times 10^{-4}$ \\
 \hline\hline
 Si XIV   & 6.180-6.186   & 6.168 & 4.915$\pm$0.471  & 3.195$\pm$0.345 \\
 Al XIII  & 7.171-7.177   & 7.155 & 4.335$\pm$0.468  & 2.819$\pm$0.336 \\ 
 Al XII r & 7.757         & 7.720 & 1.044$\pm$0.253 & 1.061$\pm$0.263 \\
 Al XII i & 7.804-7.807   & 7.780 & 0.760$\pm$0.210 & 0.772$\pm$0.217 \\
 Mg XI    & 7.850         & 7.820 & 0.516$\pm$0.166 & 0.538$\pm$0.082 \\
 Al XI f  & 7.872         & 7.855 & 0.8320$\pm$0.331 & 0.844$\pm$0.338 \\
 Mg XII   & 8.419-8.425   & 8.380 &   8.441$\pm$0.572  & 8.576$\pm$0.722 \\
 Mg XI r  & 9.169         & 9.135 & 2.233$\pm$0.402         & 9.256$\pm$1.729 \\
 Mg XI i  &  9.238-9.231  & 9.190 & 1.167$\pm$0.448  & 6.731$\pm$0.387 \\
 Mg XI f  & 9.314         & 9.285 & 2.306$\pm$0.577 & 9.421$\pm$0.255 \\
 Ne X     & 10.24         & 10.210 & 2.051$\pm$0.449 & 20.51$\pm$4.61 \\
 Ne X     & 12.132-12.138 & 12.105  & 5.801$\pm$0.965 & 871.10$\pm$151.29 \\
                \hline
                \hline
        \end{tabular}
 \end{table}
 \begin{table}
 	\small
 	\caption{Parameters in the model fits to the spectra of the HETG (September)
 		and LETG (December) \added{with their 90 \% confidence
 range uncertainties}. The abundances are given in Table 
\ref{table:abundances}.}
 	\label{table:params}
 	\centering
 	\begin{tabular}{lll}
 		\hline\hline
 		Parameter  & September  & December \\
 		\hline 
 		\hline
 		\textbf{General Parameters} &&\\
 		\hline &&\\[-0.4cm]
 		Flux(abs) erg cm$^{-2}$ s$^{-1}$       &$1.7^{+0.1}_{-0.1}\times 10^{-11}$&$8.6^{+0.3}_{-0.3}\times 10^{-12}$ \\  
 		Flux(unabs) erg cm$^{-2}$ s$^{-1}$     &$6.8^{+0.7}_{-0.2}\times 10^{-11}$ & $4.1^{+0.6}_{-0.2}\times 10^{-9}$ \\
 		N(H) $\times 10^{21}$ cm$^{-2}$         &  32.2$^{+5}_{-6}$     & 5.6$^{+2}_{-2}$  \\
 		v$_{\mathrm{broadening}}$ (km s$^{-1}$) & 676$_{-70}^{+80}$     & 1190$_{-45}^{+210}$     \\
 		\hline
 		\textbf{First component} &&\\
 		\hline &&\\[-0.4cm]
 		T$_1$ (keV)                             &  4.46$^{+0.6}_{-0.3}$   & 0.071$^{+0.03}_{0.03}$  \\ 
 		Flux$_1$(abs)  erg cm$^{-2}$ s$^{-1}$   & $1.4^{+0.1}_{-0.1}\times 10^{-11}$  & $9.4^{+3.0}_{-3.0}\times 10^{-13}$   \\
 		Flux$_1$(unabs) erg cm$^{-2}$ s$^{-1}$  & $2.9^{+0.2}_{-0.2}\times 10^{-11}$ & $4.9^{+2.0}_{-2.0}\times 10^{-10}$ \\
 		EM$_1$ ($10^{54}$ cm$^{-3}$)            & $12.1^{+0.6}_{-0.4}$    & $2.3^{+0.5}_{-0.4}$ $\times 10^{3}$ \\
 		v$_{\mathrm{blueshift}}$                & frozen 0 & frozen 0\\
 		\hline
 		\textbf{Second component} &&\\
 		\hline &&\\[-0.4cm]
 		T$_2$ (keV)                             & 0.78$^{+0.4}_{-0.4}$        & 0.49$^{+0.3}_{-0.3}$ \\
 		Flux$_2$(abs)  erg cm$^{-2}$ s$^{-1}$   & $3.1^{+0.1}_{-0.3}\times 10^{-12}$  & $3.1^{+0.3}_{-0.3}\times 10^{-12}$    \\
 		Flux$_2$(unabs) erg cm$^{-2}$ s$^{-1}$  & $3.9^{+0.1}_{-0.3}\times 10^{-11}$ & $1.8^{+0.2}_{-0.2}\times 10^{-11}$    \\    
 		EM$_2$ ($10^{54}$ cm$^{-3}$)            & $4.9^{+0.5}_{-0.2}$    & $1.6^{+0.2}_{-0.4}$  \\                         
 		v$_{\mathrm{blueshift}}$ (km s$^{-1}$)  & -856$^{+75}_{-145}$     & -270$^{+70}_{-70}$ \\
        \hline
        \textbf{Atmospheric component} &&\\
        \hline &&\\[-0.4cm]
 		T$_{\mathrm{atm}}$ (K)                               & & 680,000$^{+20000}_{-20000}$  
 		\\ v$_{\mathrm{blueshift,atm}}$ (km s$^{-1}$)        & & $2260^{+260}_{-280}$ \\
 		Flux$_{\mathrm{atm}}$(abs)  erg cm$^{-2}$ s$^{-1}$   & & $4.6^{+0.3}_{-0.3}\times 10^{-12}$ \\
 		Flux$_{\mathrm{atm}}$(unabs)  erg cm$^{-2}$ s$^{-1}$ & & $3.5^{+0.2}_{-0.2}\times 10^{-9}$ \\
 		\hline
 		\hline
 	\end{tabular}
 \end{table}
 \begin{table}
 	\small
 	\caption{Abundances of the models fitted to the spectra of the HETG (September)
 		and LETG (December). The abundances \citep[relative to solar values][]{Asplund06}
 		are assumed equal in  the two components of thermal plasma in CIE.
 		The abundances marked with an asterisk were
 		fixed in the fit, as explained in Section 3.}
 	\label{table:abundances}
 	\centering
 	\begin{tabular}{lccc}
 		\hline\hline
 		Element & September  & December & Optical            \\
 		\hline                                                            
 		\hline                                                            
 		He/He$_\odot$  & 2.5$^*$              & 2.5$^*$             & 1.5 \\
 		C/C$_\odot$  & 0.9$^*$              & 0.9$^*$             & ... \\
 		N/N$_\odot$  & 100$^*$              & 100$^{+160}_{-35}$  & 140 \\
 		O/O$_\odot$    & 290$^*$              & 290$^{+160}_{-40}$  & 2.3 \\
 		Ne/Ne$_\odot$  & 1660$^{+40}_{-160}$  & 190$^{+80}_{-120}$  & 40  \\
 		Mg/Mg$_\odot$  & 230$_{-20}^{+40}$    & 220$^{+70}_{-60}$   & ... \\
 		Al/Al$_\odot$  & 250$^{+70}_{-70}$    & 770$^{+380}_{-380}$ & ... \\
 		Si/Si$_\odot$  & 27$^{+5}_{-6}$       & 60$^{+40}_{-40}$    & ... \\
 		S/S$_\odot$    & 26$^{+15}_{-13}$     &                     & ... \\
 		Ar/Ar$_\odot$  & 37$_{-32}^{+28}$     &                     & 0.3 \\
 		K/K$_\odot$    &  1340$^{+830}_{-830}$&                     & ... \\
 		Ca/Ca$_\odot$  & 30$_{-28}^{+30}$     & 900$^{+900}_{-600}$ & ... \\
 		Fe/Fe$_\odot$  & 4$^{+5}_{-3}$        &                     & 1.2 \\
 		\hline
 		\hline
 	\end{tabular} 
 \end{table}
 
The model's best fit parameters are in Tables \ref{table:params} and \ref{table:abundances}.
 The best fit  negative velocity in the cooler component is -856 km s$^{-1}$ and
 the lines are broadened by 760 km s$^{-1}$, much less than the  the  
 full width at half maximum (FWHM) measured
 in the optical emission lines and  attributed to the 
 expansion velocity of the two poles by \citet{Ribeiro13} and \citet{Shore13},
 \added{reaching a maximum 3000 km s$^{-1}$. }
 These groups of authors measure for most emission lines a null center velocity,
 with equal contribution of blue and red shifted material.
 First \edit2{we notice that} the flat-topped profile of the less
 hot component that mostly
 contributes to the line fluxes
 is difficult to reconcile with the bipolar outflow inferred
 from the profile of the optical lines. It seems more likely
 that in the X-ray emitting material
 there is ``only a modest distortion from spherical'', as suggested by  
\citet{Ignace06} for similarly shaped X-ray emission lines
 of hot stellar winds. Moreover, as already noted above, 
we cannot rule out that a  redshifted
 portion of the lines may have been completely
 absorbed, due to optical depth increasing
 with radius. This effect
 would  shift the apparent peak of the line bluewards and reduce its width  
 \citep{Ignace16}; the oblique profile towards the red seems to confirm
 this interpretation. 
 The blueshift of the line center may also be due to the
 average velocity of the X-ray emitting medium, while the broadening may 
 be caused by the spread of the flow in space; with the present
data it is difficult to discriminate between the two cases.
  
 In order to make our assumptions as consistent as possible \edit2{in each of the} two {\sl Chandra} observations
 and \edit2{maintain consistency with} the optical \edit2{results} as well, we assume 
 that the most abundant elements, He, C, N, and O have the same abundances throughout observations,
 \edit2{and that} clumping and ``late dredge-up effects'' did not change  abundances
 significantly while the outburst was still in process and the nova  ejecta evolved.
 Therefore, we fixed the He abundance to match 
 Tarasova's estimates,  we assumed  an approximately solar C
 as in many ONe novae (we adopted the value suggested by \citet[][]{Shore13}), 
 and  we fixed the abundances of N and O to 
 match the values we obtained in the fit to the December spectrum,
 in which prominent emission lines
 of these elements were measured.  The other abundances we derived
  are presented in Table \ref{table:abundances}.
 The relative abundances of the elements
 with atomic weight larger than oxygen
(for example the ratio of Mg or Al to Ne) depend very little on
 the normalization obtained by fixing the abundances
 of the less heavy elements, so our assumptions on the H, He, C, O and N abundances
 for this spectrum are not critical.

 \edit2{We also attempted fitting}
the data with the  cooling flow model VMCFLOW in XSPEC, in order to test
 the possibility of a smoothly varying temperature as the X-ray emitting
 plasma travels farther from the WD and cools.
  We regard this as an experiment, because
 the cooling flow model in XSPEC does not
 allow to parameterize the line broadening \edit2{and} to account for different line
 velocity other than cosmological redshift. \edit2{Moreover,} it includes abundances of
 fewer elements than the APEC models.  The best fit
 with this model returned a  $c$=6663 value for 7918 degrees of freedom, but only the strength of
 the most prominent lines is matched, while the continuum was less well
 fitted than with the apparently more simplistic
 two component model.  This fit returned lower N(H) than
 the two-component fit, namely 1.71 $\times 10^{21}$ cm$^{-2}$, 
 a maximum temperature of 4.9 keV and minimum temperature
 of 100 eV. We note that this lower value is consistent with the suggestion  
 for the {\sl Swift} XRT short observations around the same epoch \citep{Mukai14}. All the 
 abundances are remarkably lower than in the two-component model,
 by a factor from 3 to 5, 
 however because no line broadening is included, the line
 fluxes are significantly underestimated by fitting narrow lines,
  thus the returned abundances are not reliable. Nevertheless, there are
 some indications we can derive from this additional modeling
 for a consistency check. The low minimum temperature
 in the VMCFLOW model may indicate that the two-component model
 has underestimated  the average temperature of the less hot gas, in which there
 may be cooler components. For Ne, this may result in largely overestimating the
 abundance because the peak of emissivity for the strong H-like Ne line is
 at lower temperature, at 0.54 keV.  We will return to this point in Section 4. 

 The VMCFLOW model is normalized with the
rate of in-flowing mass $\dot m$ , which is a parameter of the model, and corresponds 
 to the mass loss rate for a nova. $\dot m$ in the fit turns out to be only 
 1.2 $\times 10^{-8}$ M$_\odot$ yr$^{-1}$, two orders
 of magnitude lower than the estimates by Shore and
 Tarasova.  Constraining $\dot m$ 
 a value of 10$^{-6}$ M$_\odot$ yr$^{-1}$ does not result in an
 adequate fit, but this is not necessarily  another shortcoming of this model;
 in fact we shall discuss in Section 4 that only a small
 fraction of the outflowing material must be emitting X-rays. 
 
\subsection{The spectrum observed in December 2012: the emergence of
 the supersoft source}
 Tables \ref{table:params} and \ref{table:abundances}
show also the parameters
 of a fit to the LETG spectrum observed in December of 2012
 with two plasma in CIE components and a WD atmosphere (Fig. \ref{fig:atmo}). The
 first important result is the \edit2{much lower} 
 equivalent column of absorbing hydrogen N(H). 
  Because
 the intrinsic absorption of the ejecta had decreased at this date, 
 N(H) \edit2{was reduced} to 5.6 $\times 10^{21}$ cm$^{-2}$, which is consistent with 
 the measurement towards the direction of the nova
 and \replaced{dos}{does} not imply intrinsic absorption.
 It is important to notice that
 \citet[][]{Shore13}  found that the optical spectrum observed
 in November of 2012 was consistent with photoionization 
 by a source at about 300,000 K, quite lower than the WD
 temperature we derive for December. Thus, the emergence of the supersoft
 source was not only due to the thinning column density, but to actual
 shrinking of the WD radius at constant bolometric luminosity,
 while the atmosphere became hotter, as
 foreseen by all the nova models \citep[e.g.][]{Wolf13}. 

 The fit whose parameters are shown in
 table \ref{table:params} was done with two CIE plasma components
 and a hot WD atmosphere. We were not able to obtain a more
 statistically significant fit adding to the atmospheric model 
 more components at different temperatures, or with the VMCFLOW model. 
 \edit2{Our proposed 3-component best fit indicates} that the hot component seems to have completely
 disappeared, and that the component \edit2{in which the blue-shifted lines are produced} 
 \edit2{has} cooled from 0.78 keV to 0.49 keV. The line centers in this spectrum are 
 still blueshifted, but by only 
 270 km s$^{-1}$.  The new, cool component of CIE plasma 
 reaches the XSPEC BVVAPEC model lower limit
 of $\simeq$70 eV, and fits the
  emission lines in the low energy (i.e. supersoft) band.
 We suggest that these lines are {\it not} associated
 with the WD: they are too strong to be attributed to the WD atmosphere
 \citep[compare with models by ][]{Rauch10,Vros12}, and consist of different species and
 ionization potentials than the observable WD absorption features.
Superimposed on the WD atmospheric continuum
 we measure in this spectrum  several Ca lines and  a strong N line at 29.2 \AA.  
 The main purpose of this observation was to measure the temperature of
 the central, supersoft X-ray source, i.e. the WD.  
 The atmospheric temperature seems to be well constrained at 680,000 K by fitting
 the supersoft portion
 of the spectrum with a static model in which the absorption features are blueshifted
 by 2260 km s$^{-1}$. \added{The specific model with
 which we obtained a best fit, chosen in  Rauch's model grids 
 described \citet[][]{Rauch10} and publicly available is the
 one in the table called SSS\_003\_00010-00060.bin\_0.002\_9.00.fits.}
 \added{We note that an alternative blackbody model component did not allow us to obtain
 a statistically significant fit; we froze the parameters of the BVAPEC components
 to the values obtained with the atmospheric fit, in order to  have
 fewer parameters to fit, and we obtained a 
 blackbody temperature of 30 eV (360,000 K), but the continuum level was 
 largely overestimated and the fit was much less statistically significant than the one
 with the atmospheric model}. 

The blueshift  in the absorption features implies that there was be a residual wind, possibly
 with low mass loss rate, even when the photosphere has receded to a radius 
 close to that before the outburst. It has been pointed out by \citet{Vros12} that
 in this case the static model only gives an upper limit for the
 atmospheric temperature, because the flux is still emitted by a larger surface
 than that of a static WD, although for
 the nova examined by van Rossum, V4743 Sgr, we find that assuming
 the same column density N(H) the difference in temperature \explain{sentence
 has been reworded here}
 the static model with appropriate abundances and the wind model 
 is less than 100,000 K.   Since no wind-atmospheric model is available
 for such extremely non-solar abundances as those predicted by the models 
 in the atmosphere of both CO and ONe H-burning WDs \citep[][]{Rauch10}, 
 we used the static atmosphere model as  best approximation. 
 We suggest that the best-fit temperature of the static model, 
 for this spectrum which is observed near supersoft maximum luminosity 
 \citep[see the {\sl Swift}-XRT hardness ratio curve 
 plotted by][]{Page13}, 
 returns a closely approximated value for the maximum T$_{\rm eff}$.
 Even if the static
 model does not account for the residual small expansion
 of the atmosphere, it does fit the absorption features 
 well,
 as it has been noted for
 other novae as well \citep[see also the case of V2491 \replaced{Cug}{Cyg},
[][]{Ness09}.  If this is the peak temperature,
 the T$_{\rm eff}$ indicates  a WD mass only slightly
 larger than 1.1 M$_\odot$ \citep[see][]{Wolf13}, however
 we know from \citet{Page13} that the peak of supersoft emission
 occurred about 20 days later, so the  WD may have \edit2{later} \deleted{still, and}
become hotter. 
 We used also
 for this fit C-statistics, due to the low counts/bin, and obtained $c=3194.91$
 and 2622 degrees of freedom.  
 
   \explain{Some words were deleted} The luminosity of the supersoft component 
 obtained from the fit is too low
 to originate from the whole WD surface. We find an absorbed flux of 
 \explain{all following numbers were corrected} 4.6 $\times 10^{-12}$ erg cm$^{-2}$ s$^{-1}$ and an unabsorbed flux
 of 3.5 $\times 10^{-9}$ erg cm$^{-2}$ s$^{-1}$, corresponding to a luminosity
 of $10^{36}$ erg s$^{-1}$ in the 0.15-1 keV band at 1.4 kpc distance. 
 At the derived 
 atmospheric temperature, the X-ray luminosity is only about 1.7\% of the
 bolometric luminosity,  6 $\times 10^{37}$ erg s$^{-1}$, calculated by \citet[][]{Wolf13}
 for  a WD with the effective temperature returned by the fit. 
Of course, we are measuring the average flux and luminosity over
 the whole observation, and as we explain in detail in Section 5,
 the supersoft flux is modulated with the orbital period. We can thus conclude that
on average, during the orbital period
 only \replaced{about 30 \%}{a very small portion} of the WD surface is observed. We suggest this may
 be due to a non-disrupted, or re-formed, accretion disk, because of
 the high inclination and the consequent orbital modulation. During the
 orbital period different portions 
 of the WD are visible, but largest portion of the WD 
 surface is obscured by the accretion disk. Section \ref{sec:timing} presents the
 resulting light curve modulation, already observed by \citet{Page13}.

	\label{fig:mgal}
%
\section{The mass of the X-ray emitting material and the chemical yields}
 The fit is normalized using the emission measure
 EM=$\int n_e n_{\mathrm i} \mathrm d V\approx n_e n_{\mathrm i} V$,
 where $n_e$ and $n_{\mathrm i}$ are the electron and ion densities and
 $V$ is the total emitting volume.
The mass M of the material in which the X-rays originate
 is dominated by the ions, $M=Y m_p N$, where m$_p$ is the
 proton mass and  $Y m_p$ is the mean atomic weight of the $N$ ions.
In such a metal rich plasma, one cannot consider the mass
 of hydrogen atoms as representative; in fact  \citet{Tarasova14} found
 the ejecta to consist of about 53\% H and 33\% He. We assume
 here that the chemical composition of the optically-emitting
 and the X-ray emitting plasma is approximately the same.
 Moreover, $N=X~n_H~V$ or $N=X~EM/n_e$ where $X$ is the mean proton number, 
$X=1.5$ and $Y=7.5$. The value of Y is dominated by
the atomic weight of
 O (38\%) and Ne (30\%), with H and He contributing  25\%.
The mass of the X-ray emitting plasma is  \begin{equation}
 M=X Y~m_p~\frac {EM}{n_e}\approx11.25~m_p~\frac {EM}{n_e}
 \end{equation}
and is crucially dependent on the value of the electron density $n_e$.

  The most sensitive diagnostic of electron density is the  R=$f/i$ line ratio, where $f$ and $i$ are the
 forbidden and intercombination line fluxes of an He like triplet
 \citep{Gabriel69, Porquet01}.
In fact, when this ratio is low, the density
 is so high that the collisional de-excitation rates are close
to the radiative decay rates. When the two rates are comparable,
the radiative decay fraction from an excited level is reduced and so
the emission becomes weaker. \citet{Orio13} found that  
 the Ne IX He-like triplet in the U Sco X-ray grating spectra
 was not consistent with the low density limit.
 For the September spectrum,
 the value that we obtain for the Mg
 He-like triplet measured in Table \ref{table:fluxes}
 is R=1.40$\pm$0.52. With R$<$1.92,
 the 1 $\sigma$ confidence maximum value, 
 the above authors found an electron density $n_e>6 \times 10^{10}$ cm$^{-3}$
 unless there is a very
 strong source of photoexcitation. We note that the density
 is at least of the order of 10$^{9}$ cm$^{-3}$ within the
 3 $\sigma$ error of the ratio we measured. The \edit2{effective temperature
 of the ionizing central source was} about 300,000 K two months later 
\citep[][]{Shore13}. Even \edit2{with the WD at this effective temperature (but almost
 certainly it was still less hot)},  all the
 emitting plasma should have been within about 6 $\times 10^{10}$ cm
 from the WD to undergo photoexcitation. 
Such a distance is quite interior in the ejecta,   
 but with {\sl Swift} an approximately constant level of 
 X-ray flux was observed since 4 weeks before the HETG exposure.
 Because we are measuring a velocity of hundreds of km s$^{-1}$,
 it is not conceivable that the X-ray emitting plasma
 is confined \edit2{in} this a small volume. 

  It thus seems
 that  in September there was not sufficient ultraviolet flux to alter
 the R value, so  high electron density seems to be the
 likely conclusion \citep{Gabriel69, Porquet01}:  
 its value in September must have 
 been indeed high enough to quench the forbidden line.
 
   As mentioned in Section 2,
the value of n$_e$ found by Shore and by Tarasova to be consistent with
 the optical spectra is  
about 10$^7$ cm$^{-3}$. This much higher  n$_e$ 
 derived from the X-ray HETG spectrum implies that the X-ray emission
 comes from a very small fraction of the ejecta, not more than 
 one hundredth the mass of the optically emitting shell;  most likely
 the ejected X-ray emitting material was in very dense
 clumps. The sum of the emission measures of the two plasma components
 for the September spectrum is 6.1 $\times 10^{55}$, implying 
 M=1.15 $\times 10^{33}$/$n_e$, or that the X-ray emitting mass M 
  was \begin{equation} M<5.75\times10^{-10}~M_\odot
 \label{eq:mass} \end{equation}
 assuming  $n_e>10^{9}$ cm$^{-3}$. This is only a small fraction
 of the mass of
  1.8 $\times 10^{-5}$ $M_\odot$, derived above from Tarasova's work,
 after correction for the revised distance, for the ejecta emitting at optical
 wavelengths.
 
  In the December LETG spectrum we do not have density diagnostics
 to assess whether the emission of the material emitting in CIE conditions
 was still very dense and clumpy. We do not rule out that the X-ray emitting
 mass had actually increased, as more mass was ejected, and
 that the lower X-ray flux is due to a much more diffuse, as opposed
 to less massive, X-ray emitting material. 
 
 For the September HETG spectrum, measured with better signal to noise
 than the December one because the nova was more luminous at the earlier date in the
 range of the strongest X-ray emission lines,
 the line fluxes in Table \ref{table:fluxes}
 can be used for each species to compare the
 yield in the X-ray emitting material with the abundances predicted by
 the global model. The global
 model fit over- or under-estimates the flux
 of the single emission lines, but gives a statistically significant
 overall fit. By making assumptions only on two parameters
for which the uncertainty is not very large, 
 namely the plasma temperatures and
 column density derived in the global fit, we found the tabulated
 emissivities for each line (N(H) must be assumed because the emissivity
 is based on the unabsorbed flux). 

 The emissivity of a line emitted at temperature T$_{\mathrm e}$ is
 defined by the following equation: 
\begin{equation}
 \mathrm{Flux} [\mathrm{ph~cm}^2 ~ \mathrm s^{-1}] = \frac{\epsilon(T_{\mathrm e})Z_{\odot}}{4 \pi d^2} \int n_{\mathrm e} n_{\mathrm {Z}} dV
\end{equation}
where the integral represents the
 emission measure for a so called ``Z'' elements, of which $n_Z$ ions are present. $Z_{\odot}$
 is the solar abundance of the given element. 
 The yield of a given ion, Y(Z) is 
\begin{equation}
 {\mathrm {Y(Z)}} =  {\mathrm m(Z)}  \frac{ \int n_{\mathrm e} n_{\mathrm Z} dV}
{ n_{\mathrm e}}\end{equation}
 where m(Z) is the mass of the single ion, obtained 
 by multiplying  its atomic weight by $m_p$.

  We obtained the tabulated values for the emissivity from the ATOMDB database
 and used the value of the flux to substitute the integral of equation
 3 into  
 equation 4. In this way, we calculated the yield of each 
 element divided by $n_e$, which in turn was divided by
 equation 1 in order to compute the abundance of the given ion,
 {\it independently of} $n_e$ (because this is also a factor in equation 1). 
We can reasonably assume that most of the elements are in the form
 of H-like and He-like ions in the plasma at 0.78 keV and in the form
 of H-like ions for the plasma at 4.5 keV. We found the emissivity of the relevant lines
 at the given temperatures, and  summed the yields
 obtained from the flux of  the H-like and He-like ions (we neglected the latter for Ne).
 As a first approximation, for the H-like lines we assumed that the mass
 fraction of the hot and less hot plasma components are given by the ratios
 of the emissivities computed for the two components in the model fit in Table
 \ref{table:fluxes}. 

 With this simple calculation,
 we obtained an independent check of the abundances
 of three
 elements of interest, Ne, Mg and Al. By adding the results of the H-like and He-like r line,
 as representative of the whole abundance of these elements, we obtained
 the following values:  N/Ne$_\odot$=436$\pm$126,  Mg/Mg$_\odot$=229$\pm$89,
Al/Al$_\odot$=170$\pm$49 (translating into the following abundances
 by number: N/Ne$_\odot$=587$\pm$169, Mg/Mg$_\odot$=308$\pm$146, 
Al/Al$_\odot$=229$\pm$66) (note that here we neglected the 
error \edit2{due the uncertainty} in the distance).
 When we compare these values with the values in Table \ref{table:abundances}, respectively 1660$^{+40}_{-160}$,
 230$^{+40}_{-20}$, and 250$\pm$70, we find  a discrepancy for the Ne abundance.
   The fact that both Al and Mg
are consistent with our model indicates the global  fit
 with two components  is compensating part of the continuum flux
 with flux in the Ne line, so that the 
 Ne abundance, while probably high, is lower.  For both Ne and Mg abundances,
 which are very sensitive to the content of plasma at temperature
 in the 0.3-0.8 keV range, the model's constraints are
 not very stringent and the uncertainty  
 is still larger than than the statistical uncertainty of the best fit we chose.
 Because of the low S/N and few counts per bin, 
 a more detailed fitting is not possible with our data, but more
 accurate determinations of these abundances will be possible in the future 
 for ONe novae observed during the phase of strong emission lines
 in the range below 20 \AA \ with better S/N. This should be done
already with the {\sl Chandra} HETG in new cases \replaced{iof}{of}
 ONe novae effected by lower
 column density N(H), or larger X-ray emitting mass and total X-ray flux.  
 In the future, {\sl Athena}
will make such measurements possible for a majority of ONe novae.

 The abundances by number of the most  abundant elements ejected
 in the ISM by other ONe novae, obtained from the analysis of ultraviolet and
 optical spectra of these novae with the CLOUDY code, are shown
 in Table \ref{table:abundances-other}, together with
 Tarasova's result for Ne and our results.
 Despite the still large uncertainties, it seems clear that the X-ray
 emitting ejecta of V959 Mon were richer in Ne, Mg and Al than the
 UV-emitting ejecta of other ONe novae. This probably implies
a high WD mass; in fact we remind that
in the models, the abundances of the elements emitted in the ISM increase
 with the WD mass,  although the same is often not true for their
 total yield, because the ejected mass decreases
 with WD mass \citep[in fact the pressure at the base of the envelope to eject mass is
reached earlier, see][]{Bennett13}.

\subsection{The detection of Al lines} 
 A very important aspect of the X-ray spectra of V959 Mon is the detection of
 Al lines. The lines of this element have previously been observed
 only in the ultraviolet
 spectra of novae. For the September spectrum, we were able 
 to determine that Al/Al$_\odot>$70, and most likely its value is in the 170-250 range. 

 A diffuse Galactic gamma-ray emission \deleted{line}
line at 1809 keV indicates radioactive
decay of $^{26}$Al into $^{26}$Mg, due to a steady amount of
 2.7$\pm$0.7 M$_\odot$ of  this isotope in the Galaxy, explained with ongoing
nucleosynthesis \citep[][]{Mahoney82, Diehl95}.
  This interesting fact may be very important for the origin
 of habitable planets, because the radioactive decay of  $^{26}$Al is
 the most effective way of heating planetesimal in proto-planetary disks
 causing differentiation and water sublimation \citep[][]{Srinivasan99}.
 In fact, $^{26}$Mg is found
 in meteorites and in presolar dust grains, implying
 injections of  $^{26}$Al in the solar system
 nebula \citep[][]{Hoppe94, MacPherson95, Huss97}. There are three
 astrophysical sites for  $^{26}$Al  nucleosynthesis: the first with most
 copious production are massive stars; the second, first
discovered by \citet{Weiss90}, are classical and recurrent novae; and the third
 are asymptotic giant branch stars
 \citep[see discussion and additional references in][]{Iliadis11,
 Bennett13}.
 If the production of novae and asymptotic giant branch stars can
 be constrained, the  rate of production of $^{26}$Al can be used to derive
 the Galactic rate of core collapse  supernovae,
 and with it the production rate of many other elements.
 By  calculating new, relevant reaction rates and
 ONe novae evolution models, \citet{Bennett13} recently came to the conclusion
 that the nova share in $^{26}$Al production amounts to 30\%.

 The fraction of $^{26}$Al  is expected to be between 11\% and 16\% of the
 total yield of Al, and increases with WD mass.  Before this work, the abundance of Al
 has been evaluated using lines in the ultraviolet
 spectrum, especially Al III $\lambda$1860, and Al II $\lambda$2669 \citep[e.g.][]{Shore03}.
  Also, the Mg abundance can be estimated from the Mg II $\lambda$2800
line in the ultraviolet, but it is
 not measured in optical spectra. We found no flux measurements
 of Al lines in the literature. Estimating
 the Al abundance from the ultraviolet
 spectra is reddening-dependent, model-dependent and has been
 particularly challenging. In a sample of five novae (see Table
 \ref{table:abundances-other} and references therein)
 only three constraining values of Al abundances have been evaluated over
 the years
 using UV spectra. The lines appeared to be produced by
photoionization by the central source, and the abundances were
 estimated by line profile fitting
 performed with the CLOUDY photoionization code \citep{Ferland13}.

 The abundance of Al by mass we obtained for the September observation with
 the flux method, 
 Al/Al$_{\odot}$=170$\pm$49  is independent of the value of $n_e$,
 although it relies on other parameters obtained from the 
 two component model.  If we assume that this abundance is representative
 of all the ejecta, including the much larger portion that emits only
 at optical wavelengths \added{(we will 
 discuss this assumption in the Discussion section)}, and that the total ejected mass was
 1.35 $\times 10^{-5}$ M$_\odot$ as referenced above, our result translates into 
 a yield of Al of
 1.3 $\times 10^{-7}$ M$_\odot$ injected in the interstellar 
 medium.  The mass fraction of Al ($^{26}$Al+$^{27}$Al) relative to
 the solar value in the ejecta is 9.3 $\times 10^{-3}$ \citep[assuming
 a solar abundance of Al by mass of 5.5 10$^{-5}$, or 2.8 10$^{-6}$ by number,]
 []{Asplund06}.  This can be compared
 with the values predicted by \citet[][]{Bennett13}, 9.5  $\times 10 ^{-3}$
 for a WD of 1.15 M$_\odot$, and 5.1 $\times 10 ^{-3}$ for
 a WD of 1.25  M$_\odot$. 
\citet[][]{Bennett13} show that, by ejecting mass fractions of Al of this order
 of magnitude, ONe novae
 contribute 30\% of the galactic yield of $^{26}$Al. 
 This  result is strongly supported by the present analysis of the V959 Mon X-ray
 data. 

\section{Timing analysis}\label{sec:timing}

\begin{figure}[t]
	\centering
	\includegraphics[]{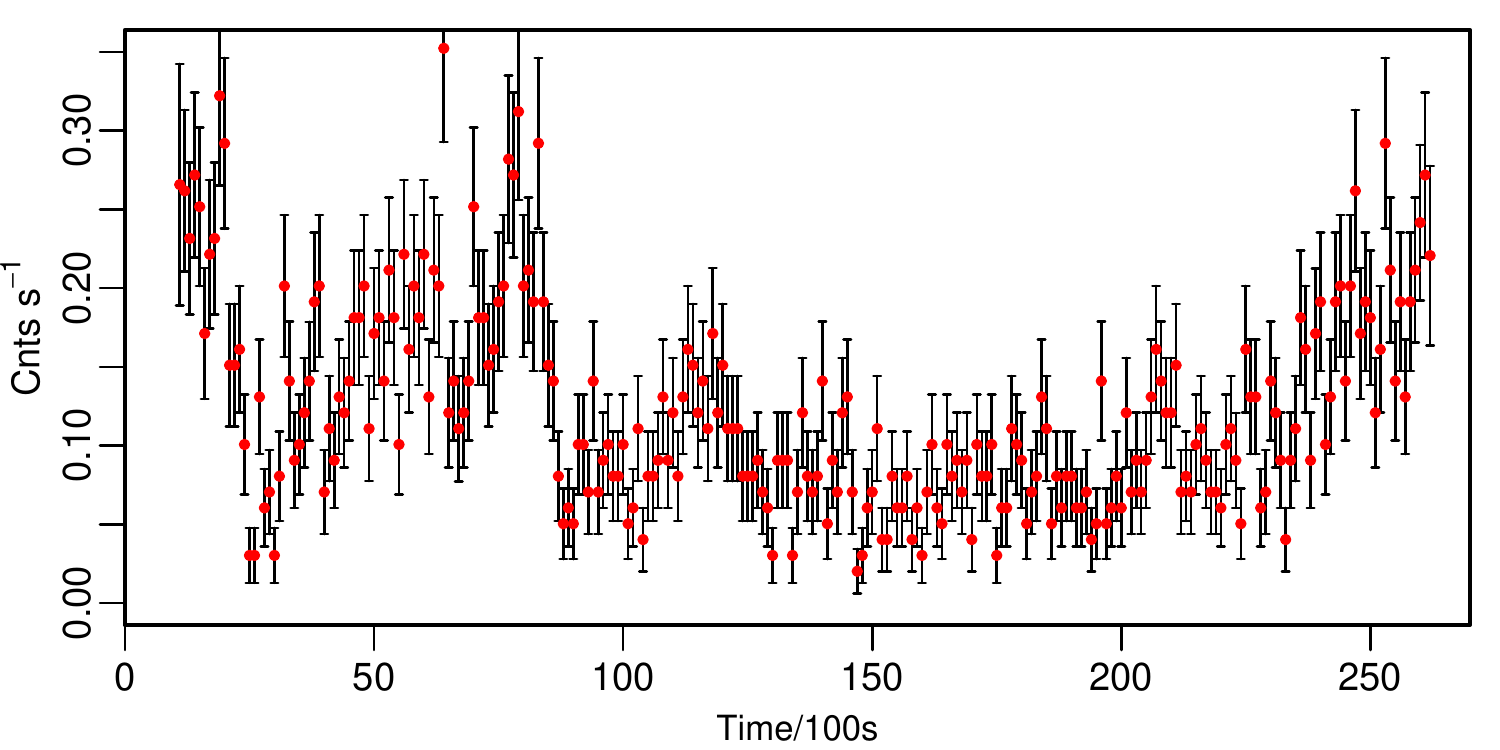}
	\caption{The light curve observed in December of 2012 with the HRC camera,
		binned every 100 s.}
	\label{fig:lc}
\end{figure}
\begin{figure}[t]
	\centering
	\includegraphics[trim=8cm 3.5cm 8cm 3.5cm,clip,width=\linewidth]{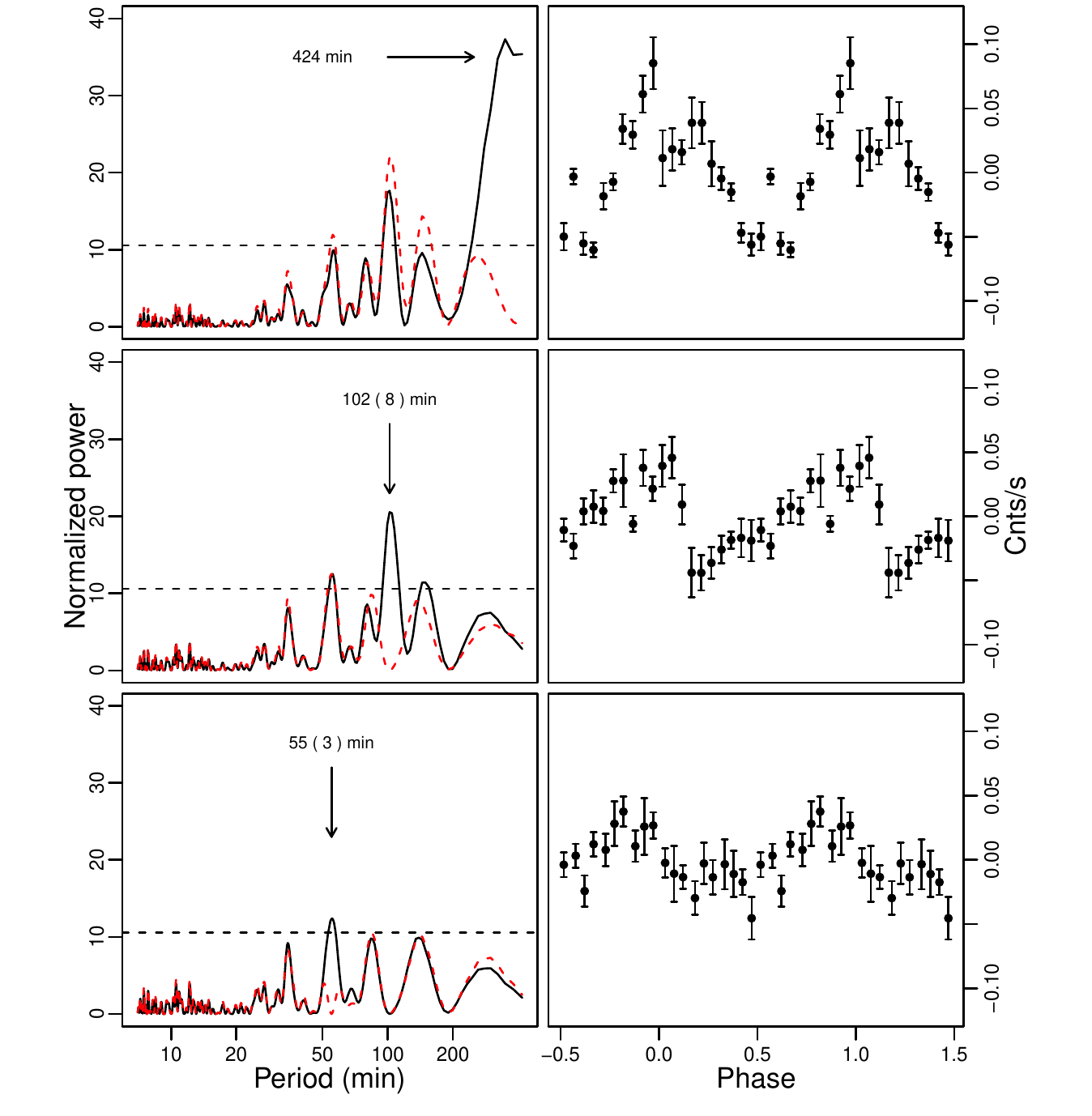}
	\caption{
		The top left panel is the Lomb-Scargle Periodogram (LSP) for the 
{\sl Chandra} HRC-S
		zero order light curve observed in December of 2012. The red dashed line shows the LSP after the subtraction of the peak corresponding to the orbital period. The middle left panel 
		shows the LSP subtracting the orbital peak and the red dashed line
		is the LSP of the data after the subtraction of the  102 min. peak. In the bottom left
		panel we subtracted the other two modulations and the red dashed line shows
		the LSP after subtracting the 55 min. period. The panels on the right, from
		top to bottom show the light curve folded with the
		orbital, 102 min. and 55 min periods respectively, with the other two modulations subtracted.}
	\label{fig:timing}
\end{figure}

While  we did not find any significant variability in the
September observation, except possibly some stochastic flickering \citep[see also ][]{Ness12} the supersoft X-ray source observed in December has
well defined modulations (see Fig. \ref{fig:lc}).
The 0.295 d (7.08 hrs) modulation reported by \citet[][]{Page13} 
 is quite prominent in our observation, although the duration of the observation
 was a little shorter than the period. We tried to subtract the 
orbital wave using the period reported by \citet[][]{Page13} and 
performed the timing analysis with the
Lomb-Scargle method. We found several peaks exceeding the height of the 0.3\% false
alarm probability level (see Fig. \ref{fig:timing}). 

 The top left panel is the Lomb-Scargle Periodogram (LSP) for the
original data. The red dashed line shows the LSP after the subtraction of the 7.08 hrs 
(orbital) modulation. The middle panel is the LSP of the data after
 subtraction of the orbital modulation and the dashed red line
is the LSP of the data after the subtraction of the highest peak, of  102 minutes.
 In the bottom
figure we subtracted both the orbital and the 102 min modulation,
 and show that a 55 minutes period emerged.
The 102 and 55 min periods are probably not entirely stable and
 may be quasi-periodic rather than periodic. 
The right panels show the phase folded light curves with the relative periods, 
after the subtraction of the other two modulations:  the top right panel is the
light curve after the subtraction of the modulations due to the 55 and 102 min
periods and folded with the orbital one, the middle left plot is the light curve
after the subtraction of the orbital and the  55 minutes modulation,
folded with the 102 min period. The bottom panel shows the light curve after
 subtracting the
 orbital and 102 min modulation, folded with the 55 min period.

All the periodic modulations are significant at wavelength below  20 \AA \
 and are only related to the supersoft component, of which the WD is
 the main source. Periods or quasi-periods of the order of an hour have already
 been observed in several novae in the supersoft X-ray source
 phase \citep[see ][]{Orio12,Nelson12} and they have been attributed to 
 non-radial g-mode oscillations of the WD.  

\section{Discussion and conclusions}

 The {\sl Chandra} spectra of V959 Mon give insight into the
 nature of the central source and the physical conditions and
 chemical composition of the ejecta.
 The observation of the emission lines due to hot
 ejected plasma in the December 2012 spectrum prompted us to examine
 the September 2012 archival spectrum. At the earlier epoch, the
 X-ray high resolution spectrum was still
 much richer in emission lines due to high ionization potential transitions
 observed with better S/N than later in December, without superimposition with
 the WD atmospheric continuum. Both {\sl Chandra} 
 spectra highlight the potential of X-ray high resolution spectra to derive
 important parameters of the nova physics.

 With some assumptions, namely that the optically
 measured abundances of He and C were the same in the plasma observed
 in the X-rays, and postulating
 that the O and N abundances were approximately constant in the two epochs
 of {\sl Chandra} observations,
 we were able to estimate absolute abundances of the special elements
 of which a nova on an ONe WD enriches the interstellar medium.
 The very large Al abundance of the two-component fit, confirmed by the yield calculation
 made by evaluating the flux in the H-like and He-like emission lines, implies
that ONe novae significantly contribute to the Galactic Al abundance,
 assuming, of course, that the abundances derived in the small X-ray emitting
 mass are representative of all the ejecta (there is no reason to
 think they may be very different).  Our results are consistent
 with the expectations that novae on ONe WDs contribute {\it at least} 30\%
 of the total galactic yield of Al$^{26}$ in the ISM.
 Future observations of the spectral region of the Al lines in these novae will 
 clarify the origin of the Galactic Al,  a problem that
 bears implications even in order to evaluate the SN II rate. 
 In the future {\it Athena} will allow
 a very accurate estimation of line fluxes in this region for a number of ONe novae.

 The lines' ratios in the September spectrum, based on the Mg XI He-like
 triplet, suggest that the X-rays were emitted by concentration
of material of very small total mass (less than 10$^{-10}$ M$_\odot$),
 concentrated  in very small clumps.  
 The spectral shape of the lines suggest a spread in space rather than  collimation.
 The remnant shell of the recurrent nova T Pyx from outbursts previous to 2011
 presents a very knotty structure that was flash-ionized again in 2011 
 \citep[][]{Shara15, Shara97}. \citet[][]{Toraskar13} attribute
 these knots to the collision of new ejecta with
 the swept-up, cold, dense shell from the older outbursts, which
 would drive Richtmyer-Meshkov instabilities in it.  
 An alternative explanation may be that a fraction of the ejecta,
 moving more slowly than the rest, is {\it from
 the beginning of the outburst} ejected in dense X-ray
emitting clumps, which later cool and constitute the clumpy material
 that was detected by \citet[][]{Shara97} with HST.
 This is only a working hypothesis, to be explored in further
research, but the concept of small clumps in the ejecta clearly
 brings to one's mind the image of the old shell of T Pyx. 

\added{\citet[][]{Williams13} attributed hard X-rays to dense blobs of gas
 colliding with each other while expanding. In this author's scenario, the
 WD ejecta are all in such blobs or clumps, and they cools very
 quickly. The more homogeneous circumbinary gas is due instead  to mass
 loss from the secondary. The small mass of X-ray emitting material at
 a given time would thus depend on the mass loss rate
 from the WD and the cooling time of the globules. 
However, for an X-ray emitting
 mass of the order of 10$^{-8}$ M$_\odot$ and a mass loss rate of the order
 of 10$^{-6}$ M$_\odot$ year$^{-1}$, a value that seem reasonable for this nova,  
 the cooling/expanding time must have been of the order of few days,
 much longer than the minute time scale predicted by \citet[][]{Williams13}. 
 This implies that \citet[][]{Williams13} model may be viable after
 carefully revisiting and constraining the physical parameters.
Incidentally, we note that, if only a portion of the material
 is ejected by the WD and a significant \edit2{part of it} is instead 
 ejected by the secondary, this makes difficult to estimate
 the yield of chemical elements from observations at any wavelengths,
 so it would be crucial to devise  ways to estimate the amount
 of ejected material from the secondary. The comparison between
 abundances of the same elements obtained in the X-ray range and in the optical range
 may be the best way to obtain this information. 
}

In the September spectrum the emission clearly originated
 in at least two regions with  distinct plasma temperatures, 4.5 and 0.78 keV,
 both in CIE. However, we could not rule out a smooth variation
 of temperature between the two values, or a lower minimum plasma temperature.
  In our best fit model, 57\% of the unabsorbed flux (but only
 18\% of the measured flux) was emitted in blueshifted material that produces
 slightly asymmetric, flat topped
 emission lines, probably indicating a large spread in spatial directions and
 radius-dependent absorption.
 43\% of the unabsorbed flux (and 82\% of the measured, absorbed flux)
originated in a much hotter region, contributing to almost a third of the flux
 in the H-like emission lines.
 The discrepancy in the evaluation of the Ne  abundance
 obtained with two methods, the global fit and the measurement
of  the flux in the lines (assuming only
 the N(H) and emission measure values estimated with the global fit),
 may imply that the temperature structure is more complex and/or
 more smoothly varying than our two zones model. 
 In the future, we should study a cooling flow type of model,  with gradually
 changing emissivity, including line broadening and other elements
 appropriate for the nova physics. In addition to working on the models, we also
 hope to obtain better quality data for  future ONe nova outbursts.

Only one component of plasma in CIE  with kT around half a keV 
 was found for the December spectrum,
 while new emission lines, that most likely are not associated with
 the central source, had emerged, probably indicating a much
 cooler plasma (kT$\simeq$0.1 keV or less), at the lower limit of the temperature
 of the CIE plasma models in XSPEC or other analysis packages.
 
The main goal we had in mind in proposing the December 2012 LETG
 observation was to measure the WD temperature and chemical composition.
 It is quite remarkable that 
 the hot WD, emitting supersoft X-rays, emerged and became
 detectable even at high inclination and with a considerable amount of 
 interstellar absorption towards the nova. The modulation
 of the central supersoft X-ray source confirms the high inclination
 at which the system is viewed. 
 This bears implications \added{for those cases in which
 the central supersoft X-ray source (SSS) is never detected.  
\citet[][]{Ness13} have suggested that the detection of the SSS 
 may be impossible in nova system viewed at high inclination,
 while in these novae the emission line spectrum of
 the ejecta is better observed and more easily measurable. As a consequence, 
 unless the \edit2{central source} detection
 is aided by a phenomenon of Thomson scattering \citep[see][]{Orio13},
we would not be able to draw any conclusion from ``missing'' SSS (e.g.
 whether it has not turned on yet, or
 it has already turned off) because in most systems  at high
inclination, because the SSS will never appear. Not
 knowing the range of inclinations, it would be 
 impossible to obtain statistics from the
 study of extragalactic novae like the one of \citet[][]{Henze14}.
The example of V959 Mon, where the SSS eventually
 was measurable even at high inclination and without
 clear evidence of Thomson scattering, indicates that
 a missing SSS is instead more likely
 to be due to slow evolution in an ejected
 shell that is optically thick to supersoft
 X-rays, rather than to the inclination. 
 There are several examples in the literature of novae
 that in which an emission lines X-ray spectrum later
 became an SSS X-ray
 spectrum when the SSS really emerged, like N LMC 2009,\citep[][]{Orio13b},
 so we suggest that the inclination is unlikely to be the determining factor in the SSS detection.
} 

 We find that the WD temperature was consistent with a WD of at least 1.1 M$_\odot$,
 comparing our best-fit temperature T$_{\mathrm eff}$=680,000 K
 with the current theoretical models \citep{Wolf13}.
 However, the increase in supersoft luminosity in the following two weeks
 is consistent with a further increase in T$_{\rm eff}$ \citep[][]{Page13},
 probably indicating an even higher mass. In fact,
 a blueshift of 2260 km s$^{-1}$ in the
 emission lines indicates residual mass loss, \edit2{implying that} 
 we did not observe the WD after all mass loss had ceased; the WD had
 not returned to its pre-outburst dimensions yet. However, 
 the detected absorption features are indeed in
 good agreement with those in the model for a 
static atmosphere on a WD of mass of at least 1.1 M$_\odot$ 
 (that is, a H-burning WD with peak temperature of, or above, 680,000 K). 

The detection of two periods of 55 and 102 minutes respectively in the LETG 
 light curve in 2012 December is consistent with several periodicities measured 
 in the supersoft X-ray source after a nova outburst \citep[][]{Orio12, Ness12}.  

Our final
 conclusion is that the two {\sl Chandra} grating spectra of V959 Mon highlight the potential of X-ray
 grating observations to gain insight into the nova physics, \edit2{by} deriving the
 effective temperature and relative mass of the central WD, 
 \edit2{by} better understanding the dynamics of the mass ejection, and \edit2{by} estimating
 the chemical yields in the interstellar medium.

\begin{table}[t]
	\centering
	\caption{Chemical abundances, by number, derived for ONe novae from
		optical and ultraviolet spectra, compared with the results for V959 Mon.
}
	\label{table:abundances-other}
	\begin{tabular}{lllll}
		\hline
		Nova      & Ne/Ne$_\odot$&Mg/Mg$_\odot$& Al/Al$_\odot$ & Reference       \\
		\hline
		V382 Vel  & 17$\pm$3     & 2.6$\pm$0.1 & 21$\pm$2      & \citet{Shore03} \\
		QU Vul    & $21.7\pm1.7$ & $10\pm5.1$  & 53.3$\pm$14.7 & \citet{2002S}   \\
		LMC 1990  & 62$\pm44$    & 16$\pm$7.5  & 257$\pm98$    & \citet{1999V}   \\
		V693 Cra  & 247$\pm144$  & 7.9$\pm7$   & 60$\pm59$     & \citet{1997V}   \\
		V838 Her  & $52.5\pm2.3$ & $1.4\pm0.8$ & 29$\pm$21     & \citet{2007S}   \\
		V1974 Cyg & 41.5$\pm$17  & 4.6$\pm$3   &  ...          & \citet{2005V}   \\
                V959 Mon  & 1660$^{+40}_{-160}$ & 230$_{-20}^{+40}$    & 230$^{+70}_{-70}$           & 2012 September (two-component fit) \\
                          & 587$\pm$169  & 308$\pm$146  & 229$\pm$66   & 2012 September (flux method) \\
                          & 190$^{+80}_{-120}$ & 220$^{+70}_{-60}$ & 770$^{+380}_{-380}$   & 2012 December \\
                          & 95   & ... & ...   & 2013 March-April \citet{Tarasova14} \\ 
		\hline
	\end{tabular}
\end{table}

\bibliographystyle{apj}
\bibliography{biblio}

\acknowledgments

\vspace{5mm}
\facilities{Chandra: HETG, ACIS-I, LETG, HRC-S}

\software{CIAO, HEASOFT, XSPEC}



\listofchanges
\end{document}